\documentstyle[12pt]{article}
\def\beq{\begin{equation}}
\def\eeq{\end{equation}}
\def\bea{\begin{eqnarray}}
\def\eea{\end{eqnarray}}
\def\nn{\nonumber}
\def\ba{\begin{array}}
\def\ea{\end{array}}

\def\v{\vert}

\def\r{\rangle}

\def\one{1\hskip -1mm{\rm l}}

\begin{document}
% \begin{flushright} hep-th/9701097
% \end{flushright}
%\rightline{TIFR/TH/97-01}
%\rightline{January 1997}
\baselineskip16pt
\smallskip
\begin{center}
{\large \bf \sf
       Quantum integrability of bosonic\\
       Massive Thirring model in continuum }

\vspace{1.3cm}

{\sf Tanaya Bhattacharyya\footnote{E-mail address:
tanaya@theory.saha.ernet.in } }

\bigskip

{\em Theory Group, \\
Saha Institute of Nuclear Physics, \\
1/AF Bidhan Nagar, Kolkata 700 064, India } \\
\bigskip

\end{center}

\vspace {1.1 cm}
\baselineskip=20pt
\noindent {\bf Abstract }

By using a variant of the quantum inverse scattering method,
commutation relations between all elements of the quantum monodromy
matrix of bosonic Massive Thirring (BMT) model are obtained. Using those 
relations, the quantum integrability of BMT model is established and the 
$S$-matrix of two-body scattering between the corresponding quasi particles 
has been obtained. It is observed that for some special values of 
the coupling constant, there exists an upper bound on the number of 
quasi-particles that can form a quantum-soliton state of BMT model.
We also calculate the binding energy for a $N$-soliton state of
quantum BMT model.

\baselineskip=16pt
\vspace {.6 cm}
%\noindent PACS No. : 11.10.Lm; 11.30.-j; 02.30.Ik; 03.65.Fd 

\vspace {.1 cm}
%\noindent Keywords : Bosonic massive Thirring model; 
%Quantum inverse scattering method; Algebraic Bethe ansatz; Soliton.

\newpage

\baselineskip=22pt
\noindent \section {Introduction }
\renewcommand{\theequation}{1.{\arabic{equation}}}
\setcounter{equation}{0}

\medskip

Quantum integrable field models in 1+1 dimensions are objects of interest
due to their close connections with different areas of physics as well as
mathematics [1-10]. These integrable theories have played an important role in
understanding the basic nonperturbative aspects of physical theories
relevant in the realistic 3+1 dimensional models. 
Through quantum inverse scattering method (QISM) one can establish the
integrability property of these models and 
obtain the spectrum as well as different correlation functions of the
corresponding models [4].

          Massive Thirring model in 1+1 dimensions has been widely 
studied as a toy counterpart to low
energy QCD, since it does not include many of the complications arising in
3+1 dimensions. The study of a nonlocal massless Thirring model is relevant,
not only from a purely field theoretical point of view but also because of
its connection with the physics of strongly correlated systems in one
spatial dimension. This model describes an ensemble of
non-relativistic particles coupled through a 2-body forward-scattering
potential and displays Luttinger-liquid behaviour [11] that can play a role in
real 1-dimensional semiconductors [12]. 

Massive Thirring model in 1+1 dimensions can be treated through QISM for both 
bosonic and fermionic field operators [6]. In this article, we shall focus
our attention to bosonic Massive Thirring (BMT) model.
The classical version of BMT model is described by the Hamiltonian
\bea
&&H = \int_{-\infty}^{\infty} dx \Big[ -\frac{i}{2}
\Big\{ ( \phi_1^*\frac{\partial
\phi_1}{\partial x} - 
\frac{\partial \phi_1^*}{\partial x} \phi_1 ) 
- ( \phi_2^*\frac{\partial \phi_2}{\partial x} 
- \frac{\partial \phi_2^*}{\partial x}\phi_2 )\Big\} \nn\\
&&~~~~~~~~~~~~~~~~~-(\phi_1^*\phi_2 + \phi_2^*\phi_1) 
-4\xi \phi_1^*\phi_2^* \phi_2
\phi_1 \Big] 
\label{a1}
\eea 
with the equal time Poisson bracket (PB) relations 
\bea
&&\{\phi_1(x), \phi_1(y)\} = \{\phi_1^*(x), \phi_1^*(y) \} = 0,~~~~~
\{\phi_1(x), \phi_1^*(y)\} = -i\delta(x-y), \nn \\
&&\{\phi_2(x), \phi_2(y)\} = \{\phi_2^*(x), \phi_2^*(y) \} = 0,~~~~~
\{\phi_2(x), \phi_2^*(y)\} = -i\delta(x-y) \, .
\label{a2}
\eea
It is well known that this 
BMT model is intimately connected with the derivative nonlinear
Schr\"{o}dinger (DNLS) model. 
In fact, one can generate the Lax operator of BMT model by `fusing' two Lax
operators of DNLS model with different spectral parameters [13].
The integrability of
the classical DNLS model, possesing ultralocal PB structure, can be 
established from the fact that the corresponding 
monodromy matrix satisfies the classical Yang Baxter equation [14]. 
The quantised version of this DNLS model also preserves the integrability
property. By applying QISM, 
the quantum integrability of DNLS
model is established and the Bethe eigenstates for all conserved quantities
have been constructed [14,15]. 

In an earlier work by Kulish and Sklyanin [6], the Lax operator and the
corresponding $R$-matrix for the quantum BMT model has been given,
though the detailed calculations are not being explicitly shown.
Moreover, the quantum Yang-Baxter equation (QYBE) 
at the infinite interval limit 
and hence the corresponding commutation relation between the creation and
annihilation operators have not been studied. However, it is
evident that taking the infinite
interval limit of the monodromy matrix and corresponding 
 QYBE is necessary to get the spectrum for the quantum
version of the Hamiltonian (\ref{a1}). In this context it may be mentioned
that, by applying a variant of the QISM [3] which is directly
applicable to field theoretical models, the quantum DNLS model
has been shown to be integrable [15,16]. 
 The infinite interval limit of the corresponding QYBE 
enabled us to obtain the spectrum of all the conserved quantities including
the Hamiltonian and also the two-particle $S$-matrix.
Therefore, it is interesting to explore the integrability property of the 
quantum BMT model by using the same variant of QISM that we
applied for DNLS model. In this article our aim is to establish such 
integrability property of quantum BMT model and to obtain the 
spectrum of all conserved quantities including the Hamiltonian.

The arrangement of this article is as follows.  
In Section 2, we consider 
the classical BMT model and evaluate the PB relations 
among the various elements of
the corresponding monodromy matrix at the infinite interval limit. 
Using these PB relations, the integrability
of the classical BMT model can be established in the Liouville sense. In this
section we also derive the expressions for the classical conserved
quantities of BMT model. 
In Section 3, we construct the quantum monodromy matrix of BMT model on a
finite interval and derive the corresponding QYBE. 
In Section 4, we consider the infinite interval 
limit of QYBE and obtain the commutation relations among the various elements
of the corresponding quantum monodromy matrix. Such commutation relations 
allow us to construct 
exact eigenstates for the quantum conserved quantities of BMT model by 
using the prescription of algebraic Bethe ansatz. 
In particular we are able to obtain the spectrum for the 
quantum version of the Hamiltonian (\ref{a1}). 
Furthermore we obtain the commutation relation
between creation and annihilation operators of quasi-particles associated
with BMT model and find out the $S$-matrix of two-body scattering among such
quasi-particles. In this section we also calculate the binding energy 
for a $N$-soliton state of the quantum BMT model.
 Section 5 is the concluding section.

\vspace{1cm}

\noindent \section {Integrability of the classical Massive Thirring
 model }
\renewcommand{\theequation}{2.{\arabic{equation}}}
\setcounter{equation}{0}

\medskip

The classical version of BMT model is described by the Lax operator [6]
\bea          
&&U( x, \lambda )  =   
i  \pmatrix { \xi \{\rho_1(x)- \rho_2(x)\}
-\frac{1}{4}\{{\lambda^2} - \frac{1}{\lambda^2}\}& 
\xi\{\lambda\phi_1^*(x) - \frac{1}{\lambda}\phi_2^*(x)\}\cr 
& & & \cr
\lambda\phi_1(x)- \frac{1}{\lambda}\phi_2(x) & 
-\xi \{\rho_1(x) - \rho_2(x)\}
+\frac{1}{4}\{{\lambda^2} - \frac{1}{\lambda^2}\} } \nn \\
&& \quad ~~~~~~~~~~~~~~~~
%&&~~~~~~~~~~~~~~~~~~~~\nn \\
%&&~~~~~~~~~~~~~~~~~~~~~~~~~~~~~~~~~~~~~~~~~~~~~~~~~~
%~~~~~~~~~~~~~~~~~~~~~~~~~~~~~~~~~~~~~~~~~~~~~~~~~~~~~~~~(2.1) \nn
\label{b1}
\eea
%\addtocounter{equation}{1}
where $\rho_1(x)=\phi_1^*(x)\phi_1(x)$, $\rho_2(x)=\phi_2^*(x)
\phi_2(x)$, $\lambda$ is the spectral parameter and 
$\xi$ is the coupling constant of the theory. The bosonic
fields $\phi_1(x)$, $\phi_2(x)$ satisfy the PB relations (\ref {a2}) and 
 vanish at $\v x \v \rightarrow
\infty $ limit.
The monodromy matrix on finite and infinite 
intervals are defined as
\beq
T^{x_2}_{x_1}(\lambda) = {\cal P} \exp \int_{x_1}^{x_2} U(x,\lambda) dx 
\label{b2}
\eeq
and 
\beq
T(\lambda) =  
\lim_{\stackrel {x_2 \rightarrow + \infty} {x_1 \rightarrow -\infty}} 
e(-x_2,\lambda) \left \{
{\cal P} \exp \int_{x_1}^{x_2} U(x,\lambda) dx \right \} e(x_1,\lambda) 
\label{b3}
\eeq
respectively, where ${\cal P} $ denotes the path ordering and 
$ e(x,\lambda) 
= e^{-{\frac{i}{4}} \{\lambda^2 - \frac{1}{\lambda^2} \}
\sigma_3 x} $. 

First, we want to investigate the symmetry properties  
of the monodromy matrix (\ref{b3}). It is easy to check that, 
the Lax operator (2.1) satisfies the relations
\bea
~~~~~~~~U ( x,\lambda )^* = K U ( x,\lambda^* ) K , ~~~~~~
U ( x, -\lambda ) = K^\prime U(x,\lambda ) K^\prime \, ,
~~~~~~~~~~~~~~~~~~~~~~~~~~( 2.4 a,b ) \nn
\eea
\addtocounter {equation}{1} 
where $ K = \pmatrix { 0 & \sqrt {-\xi} \cr {1 / \sqrt{-\xi}} & 0 } $ and 
 $ K^\prime = \pmatrix { 1 & 0 \cr 0 & -1 } $. 
%satisfy $ K^2 = {K^\prime}^2 = \one $. 
By using these relations, we find  
 that the symmetries of the monodromy matrix $T(\lambda)$
(\ref{b3}) are given by
\bea
~~~~~~~~~~~~~~~T(\lambda)^* = K T( \lambda^* ) K  ,~~~~~~  T ( -\lambda ) 
= K^\prime T( \lambda )
K^\prime \, . ~~~~~~~~~~~~~~~~~~~~~~~~~~~~~~~~~(2.5 a,b ) \nn
\eea
\addtocounter{equation}{1}
Due to the relation (2.5a), 
$T(\lambda )$ can be expressed in a form
\bea
T ( \lambda ) = \pmatrix { a( \lambda ) & -\xi b^*( \lambda ) \cr
                          b( \lambda ) & a^*( \lambda ) } \, , 
\label{b6}
\eea 
where $\lambda$ is taken as a real parameter.
Moreover, by using the symmetry relation 
(2.5b), it is easy to see
 that $a(-\lambda)=a(\lambda)$ and $b(-\lambda )=-b(\lambda)$.
Therefore, it is sufficient to derive the PB relations among the elements of 
$T(\lambda)$  only for $\lambda\geq 0 $. 

Next, our aim is to calculate the classical conserved quantities of BMT model
by using the approach described in Ref 2.
From (\ref{b2}), one obtains the differential equation followed by
the monodromy matrix $T_{x_1}^{x_2}(\lambda)$ as
\bea
\frac{\partial}{\partial x_2}T_{x_1}^{x_2}(\lambda ) = U( x_2, \lambda )
T_{x_1}^{x_2}(\lambda ) \,.
\label{b7}
\eea
Now, let us decompose the monodromy matrix in the 
form
\bea
T_{x_1}^{x_2}(\lambda ) = \Big(\, 1+ W( x_2, \lambda )\, \Big)\exp Z( x_2, x_1
, \lambda )\Big(\, 1+ W( x_1, \lambda)\,\Big) \, ,
\label{b8}
\eea
where $Z( x_2, x_1, \lambda )$ is a diagonal matrix and $W(x, \lambda )$ is 
a nondiagonal one. The Lax operator of the 
classical BMT model can be expressed as $U(x, \lambda ) = 
U_d( x, \lambda ) + U_{nd}( x, \lambda )$, where $U_d(x, \lambda)$ is the
diagonal part and $U_{nd}(x, \lambda)$ is the non-diagonal part of $U(x,
\lambda)$. Using
the above expression of the Lax operator $U(x, \lambda)$ (\ref{b1}), 
the differential 
equation (\ref{b7}) can be decomposed into
\bea
~~~~~~~~~~~~~~\frac{dZ}{dx}= U_d + U_{nd}W \, , ~~~~~ 
\frac{dW}{dx} - 2U_d W - U_{nd} + WU_{nd}W = 0 \, .
~~~~~~~~~~~~~~(2.9a,b) \nn
\eea
\addtocounter{equation}{1}
The structure of the Lax operator (2.1) ensures that $W(x_2, \lambda)$ and
$Z(x_2,x_1, \lambda)$ can be written in the form
\bea
&&W( x_2, \lambda ) = -\xi w^*(x_2, \lambda )\sigma_+ + w( x_2, \lambda )\sigma_- 
\, ,
\nn \\   
&& Z( x_2, x_1, \lambda )= z(x_2,x_1,\lambda) \sigma_3 \, . \nn
\eea
Substituting eqns (\ref{b6}) and (\ref{b8}) in the expression (\ref{b3}),
and using $W(x, \lambda) \rightarrow 0$ at $\v x \v \rightarrow 
\infty $ limit,
one obtains, 
$$\ln a(\lambda) = \lim_{\stackrel{x_2 \rightarrow +\infty}{x_1
\rightarrow -\infty}} \big \{ z(x_2,x_1,\lambda) + \frac{i
\lambda^2}{4}(x_2-x_1) \big \}.$$
Substituting the 
  explicit form of $z(x_2,x_1, \lambda)$ (as obtained  
  by integrating eqn.(2.9a)) to the above expression, 
we get the following form of $\ln a(\lambda)$:
\bea
\ln a(\lambda ) = i\xi\int_{-\infty}^{+\infty} \{ \phi_1^*\phi_1 -
\phi_2^*\phi_2 \} dx + i\xi\lambda\int_{-\infty}^{+\infty}\phi_1^* w dx -
\frac{i\xi}{\lambda} \int_{-\infty}^{+\infty}\phi_2^* w dx \, .
\label{b10}
\eea

Next, we expand $w( x,\lambda )$ in inverse powers of $\lambda$ as 
$$w( x, \lambda ) = \sum_{j=0}^\infty
\frac{w_j}{\lambda^{2j+1}}\, . $$ 
Using the differential eqn.(2.9b) followed by $W(x, \lambda)$, 
the expansion coefficient 
$w_j$s can be obtained explicitly in a recursive way. 
The first few nonzero $w_j$s are given by
\bea
w_0 = -2\phi_1 \,; \,~~ \, w_2 = 4i{\phi_1}_x +
8\xi\phi_1(\phi_2^*\phi_2) + 2\phi_2 \, .\nn
\eea
Substituting 
 $w_j$s in the expression of $\ln a(\lambda)$ (\ref{b10}), one gets
$$\ln a( \lambda ) = \sum_{n=0}^{\infty} \frac{iC_n}{\lambda^{2n}} ,$$ 
where $C_n$s represent 
an infinite set of conserved quantities. The first two of them are
explicitly given by
\bea
&&C_0 = -\xi \int_{-\infty}^{+\infty}\{\,\phi_1^*\phi_1 + \phi_2^* \phi_2 
\,\} dx \, ,~~~~~~~~~~~~~~~~~~~~~~~~~~~~~~~~~~~~~~~~~~~~~~~~~~~~~~
~~~~~~(2.11a) \nn \\
&& C_1 = 4i\xi\int_{-\infty}^{+\infty}\phi_1^*{\phi_1}_x \, dx + 2\xi
\int_{-\infty}^{+\infty}\{ \phi_1^*\phi_2 + \phi_2^* \phi_1 \} dx 
+ 8\xi^2 \int_{-\infty}^{+\infty}(\phi_1^*\phi_1)
(\phi_2^*\phi_2)dx \, .
~~(2.11b) \nn
\eea
\addtocounter{equation}{1} 
Next we expand $w( x, \lambda )$ in powers of $\lambda$ as
$$w(x, \lambda ) = \sum_{j=0}
^\infty \tilde{w}_j \lambda^{2j+1}.$$ 
In a similar way as above, using (2.9b), 
the first few nonzero $\tilde{w}_j$s can be obtained as 
$$\tilde{w}_0= -2\phi_2 \, ~~~~\, \tilde{w}_2 = -4i{\phi_2}_x 
+ 8\xi(\phi_1^*\phi_1)\phi_2 + 2\phi_1 \, .$$ 
Correspondingly, eqn.(\ref{b10}) yields 
$$\ln a(\lambda)= \sum_{n=0}
^\infty i\tilde{C}_n\lambda^{2n},$$ 
where $\tilde{C}_n$s represent another 
infinite set of conserved quantities.  The first two of them are 
explicitly given by
\bea
&& \tilde{C}_0 = \xi \int_{-\infty}^{+\infty} \{\,\phi_1^*\phi_1 + 
\phi_2^* \phi_2 
\,\} dx \, ,~~~~~~~~~~~~~~~~~~~~~~~~~~~~~~~~~~~~~~~~~~~~~~~~~~~
~~~~~~~~~~~~(2.12a) \nn \\
&& \tilde{C}_1 = 4i\xi\int_{-\infty}^{+\infty}\phi_2^*{\phi_2}_x \, dx - 2\xi
\int_{-\infty}^{+\infty}\{ \phi_1^*\phi_2 + \phi_2^* \phi_1 \} dx
- 8\xi^2 \int_{-\infty}^{+\infty}(\phi_1^*\phi_1)(\phi_2^*\phi_2)dx \, .
~~(2.12b) \nn
\eea
\addtocounter{equation}{1}

Now by combining these two sets of conserved quantities, the mass,
 momentum and the Hamiltonian of classical BMT model can be
expressed in the following way:
\bea
&&N = -\frac{1}{2\xi}(C_0-\tilde{C}_0) = 
\int_{-\infty}^{+\infty} (\,\phi_1^*\phi_1 + \phi_2^* \phi_2 
\,) dx \, , \nn \\
&& P = -\frac{1}{4\xi}(C_1 + \tilde{C}_1)=
\int_{-\infty}^{+\infty} (\phi_1^*{\phi_1}_x + \phi_2^*{\phi_2}_x )
 \, dx \nn \\ 
&&H = -\frac{1}{4\xi} (C_1-\tilde{C}_1) =
\int_{-\infty}^{\infty}\big[-i(\phi_1^*{\phi_1}_x - \phi_2^*{\phi_2}_x) -
\{\phi_1^*\phi_2 + \phi_2^*\phi_1\} -4 \xi \phi_1^*\phi_2^*\phi_2\phi_1\big]\,
dx. \nn
\eea

Next, we want to derive the PB relations among the elements of 
    $T( \lambda )$ (\ref {b6}). We apply the equal time PB relations
(\ref{a2}) between the basic field variables
to evaluate the PB relations among the elements of the 
Lax operator (2.1) and find that 
\beq
\left\{ U( x, \lambda ) {\stackrel {\otimes}{,}}
 U( y, \mu ) \right\} = \left [ r( \lambda, \mu ), U( x,
\lambda ) \otimes \one + \one \otimes U( y, \mu ) \right ]\, \delta( x-y ) \, ,
\label{b13}
\eeq  
where 
\beq
r( \lambda, \mu ) = -\xi \left \{ \, t^c\sigma_3\otimes\sigma_3 + s^c (
\sigma_+\otimes\sigma_- + \sigma_-\otimes\sigma_+ ) \, \right\}
\label{b14}
\eeq
with $ t^c = \frac{\lambda^2 + \mu^2}{2( \lambda^2 - \mu^2 )} , \quad s^c =
\frac{2\lambda\mu}{\lambda^2 - \mu^2} $. Now, by using the eqns.(\ref{b13})
and (\ref{b3}), one obtains 
\beq
\left\{ T( \lambda ){\stackrel{\otimes}{,}} T( \mu ) \right\} 
= r_+ ( \lambda, \mu ) T( \lambda )
\otimes T( \mu ) - T( \lambda ) \otimes T ( \mu )r_-( \lambda, \mu ) \, ,
\label{b15}
\eeq
where 
\bea
r_\pm = -\xi \left ( t^c \sigma_3 \otimes \sigma_3 + s^c_\pm \sigma_+ 
\otimes \sigma_- + s^c_\mp \sigma_- \otimes \sigma_+ \right ) \, ,
\nn
\eea 
with $ s^c_\pm = \pm 2 i \pi \lambda^2 \delta ( \lambda^2 - \mu^2 ) $. 
%\[\lim_{L \rightarrow \infty} P\frac{e^{\pm ikL}}{k} = \pm i \pi \delta(k) \] 
By substituting the symmetric form of $T(\lambda )$  (\ref {b6}) 
to eqn.(\ref{b15}) and comparing the individual elements in both sides, 
we obtain 
\bea
&&~~~~~\{ a( \lambda ) , a( \mu ) \} = 0 \, , ~~
\{ a( \lambda ) , a^\dagger( \mu ) \} = 0 \, , ~~
\{ b( \lambda ) , b( \mu ) \} = 0 \, , ~~
~~~~~~~~~~~~~~~~\nn (2.16a,b,c) \\
&&~~~~~\{ a( \lambda ) , b( \mu ) \} = \xi \left(
\frac{\lambda^2 + \mu^2 }{\lambda^2 - \mu^2} \right) \,
a( \lambda )b( \mu ) - 2i\pi\xi\lambda^2 \, \delta( \lambda^2 - \mu^2 ) \, 
b(\lambda )a(\mu)  \, ,\nn ~~~~~~~~~( 2.16d ) \\
&&~~~~~\{ a( \lambda ) , b^*( \mu ) \} = -\xi \left(
\frac{\lambda^2 + \mu^2 }{\lambda^2 - \mu^2} \right) \,
a( \lambda )b^*( \mu ) + 2i\pi\xi\lambda^2 \, \delta( \lambda^2 - \mu^2) \,
b^*(\lambda )a( \mu ) \, , \nn ~~~(2.16e) \\
&&~~~~~\{ b( \lambda ) , b^*( \mu ) \} 
= -4i \pi \lambda^2 \, \delta( \lambda^2 -
\mu^2) \, \v {a( \lambda )} \v^2 \, . \nn ~~~~~~~~~~~~~~~~~~~~~~~~~~~~
~~~~~~~~~~~~~~(2.16f)
\eea
\addtocounter{equation}{1}

From eqn.(2.16a) it follows
that all expansion coefficients occuring in the expansions of $\ln
a(\lambda)$ will have vanishing PB relations among themselves. Hence, 
the following expressions will hold true
$$ \{C_m, C_n\} = \{\tilde{C}_m, \tilde{C}_n \} = \{C_m, \tilde{C}_n\} =0,$$
for all values of $m$ and $n$.
Since the mass, momentum and the Hamiltonian of the classical 
BMT model has been expressed in terms of the expansion coefficients $C_n$
and $\tilde{C}_n$s, all of them will have vanishing PB relations among 
themselves. Thus the 
integrability property of the classical
BMT model, described by the Hamiltonian (\ref{a1}), 
is established in the Liouville sense.

\vspace{1cm}

\noindent \section 
{Commutation relations for the quantum monodromy matrix on a finite interval}
\renewcommand{\theequation}{3.{\arabic{equation}}}
\setcounter{equation}{0}

\medskip

By using a version of QISM which is directly applicable to field models [3], 
in this section we shall
show that the quantum monodromy matrix of BMT model on a finite interval
satisfies QYBE.
The basic field operators of 
the quantum BMT model satisfy the following equal time commutation relations:
\bea
\left[ \phi_1( x ) , \phi_1( y ) \right] = 
\left[\phi_1^\dagger( x ) ,\phi_1^\dagger( y )
\right] = 0 ; ~~~~
\left[ \phi_1( x ) ,\phi_1^\dagger( y ) \right] =  \hbar \delta( x - y ) \, ,\nn \\
\left[ \phi_2( x ) , \phi_2( y ) \right] = 
\left[\phi_2^\dagger( x ) ,\phi_2^\dagger( y )
\right] = 0 ; ~~~~
\left[ \phi_2( x ) ,\phi_2^\dagger( y ) \right] =  \hbar \delta( x - y ) \, ,
\label {c2}
\eea
and the vacuum state is defined through the relations $\phi_1(x) \v 0 \r = 
~\phi_2(x) \v 0 \r = 0$.

In analogy with the classical Lax operator (\ref{b1}), we assume that the 
quantum Lax operator of BMT model is given by
\bea
{\cal U}_q( x, \lambda ) = i \pmatrix { f_1\rho_1( x )-f_2\rho_2(x) 
- \frac{\lambda^2}{4} + \frac{1}{4\lambda^2} &
\xi\lambda\phi_1^\dagger( x )-\frac{\xi}{\lambda}\phi_2^\dagger(x)  \cr
&  &  &  \cr
\lambda\phi_1(x)-\frac{1}{\lambda}\phi_2(x)  &  -g_1\rho_1( x )+g_2\rho_2(x)
 +\frac{\lambda^2}{4}-\frac{1}{4\lambda^2} } \nn \\
\quad ~~~~~~~~~~~~~~
\label{c1}
\eea
where $\rho_1(x)=\phi_1^\dagger(x)\phi_1(x),\, \, \rho_2(x)= 
\phi_2^\dagger(x)\phi_2(x)$ and $f_1, f_2, g_1, g_2$ 
are four parameters which will be determined later in this section through
QYBE.
Using the Lax operator (\ref {c1}), the quantum monodromy matrix 
on a finite interval is defined as 
\beq
{\cal T} ^{x_2}_{x_1}(\lambda) = \quad  :{\cal P} exp \int_{x_1}^{x_2} {\cal
U}_q(x,\lambda) dx : \, ,
\label {c3}
\eeq
where the symbol $::$ denotes the normal ordering of operators.  
This quantum monodromy matrix (\ref {c3}) satisfies 
a differential equation given by
\bea
\frac{\partial}{\partial x_2}{\cal T}^{x_2}_{x_1}( \lambda ) &&
= : {\cal U}_q( x_2 ,
\lambda ){\cal T}^{x_2}_{x_1}( \lambda ) : \nn\\
\quad \quad \quad 
&&= - \frac{i}{4}\{\lambda^2-\frac{1}{\lambda^2}\}
\sigma_3{\cal T}^{x_2}_{x_1}( \lambda ) +
 i\xi\lambda\phi_1^\dagger( x_2 )\sigma_+{\cal T}^{x_2}_{x_1}( \lambda ) 
-\frac{i\xi}{\lambda}\phi_2^\dagger(x_2)\sigma_+{\cal T}^{x_2}_{x_1}( \lambda
)\nn \\
&&~~~+i\lambda\sigma_-{\cal T}^{x_2}_{x_1}( \lambda )\phi_1( x_2 )
-\frac{i}{\lambda}\sigma_-{\cal T}^{x_2}_{x_1}( \lambda )\phi_2(x_2)\nn\\
&&~~~+if_1\phi_1^\dagger( x_2 )e_{11}{\cal T}^{x_2}_{x_1}( \lambda )\phi_1( x_2 )
-if_2\phi_2^\dagger( x_2 )e_{11}{\cal T}^{x_2}_{x_1}( \lambda )
\phi_2( x_2)\nn \\
&&~~~- ig_1\phi_1^\dagger( x_2 )e_{22}{\cal T}^{x_2}_{x_1}( \lambda )
\phi_1( x_2 ) 
+ig_2\phi_2^\dagger( x_2 )e_{22}{\cal T}^{x_2}_{x_1}( \lambda )
\phi_2( x_2 ) \, ,
\label{c4}
\eea  
where $e_{11} = \frac{1}{2}( 1 + \sigma_3 )$ and 
$e_{22} = \frac{1}{2}(1 - \sigma_3 ) $. 
Now, to apply QISM, 
we have to find out the differential equation satisfied by the product 
${\cal T}^{x_2}_{x_1}(\lambda)\otimes {\cal T}^{x_2}_{x_1}(\mu)$.
By using the basic commutation relations 
(\ref {c2}) and the method of `extension' [3],
we find that the product of two monodromy matrices 
satisfies the following differential equation (detail 
calculations are given in Appendix A): 
\beq
\frac{\partial}{\partial x_2} 
\left({\cal T}^{x_2}_{x_1}(\lambda)\otimes {\cal T}^{x_2}_{x_1}(\mu) \right) 
= \vdots {\cal L}( x_2; \lambda, \mu )
{\cal T}^{x_2}_{x_1}( \lambda )\otimes{\cal T}^{x_2}_{x_1}( \mu ) \vdots ~,
\label{c5}
\eeq
where
\beq
{\cal L}( x; \lambda, \mu ) 
= {\cal U}_q( x, \lambda ) \otimes \one + \one \otimes
{\cal U}_q( x, \mu ) + {\cal L}_\triangle(x; \lambda, \mu ) \, ,
\label{c6}
\eeq
with
$$
{\cal L}_\triangle(x; \lambda, \mu ) = \pmatrix {-\hbar f_1^2\rho_1(x) &
- \hbar\xi \mu f_1\phi_1^\dagger(x) 
 & 0 & 0 \cr
-\hbar f_2^2\rho_2(x) & -\frac{\hbar\xi}{\mu}f_2\phi_2^\dagger(x) & & \cr
\quad & \quad & \quad & \quad \cr
0 & \hbar g_1f_1\rho_1(x) & 0 & 0 \cr
&+ \hbar g_2 f_2\rho_2(x) & &  \cr
\quad & \quad & \quad & \quad \cr
-\hbar\lambda f_1\phi_1(x)&
-\hbar\xi\{\lambda\mu +\frac{1}{\lambda\mu}\} & \hbar g_1f_1\rho_1(x) &
\hbar\xi \mu g_1\phi_1^\dagger(x) \cr 
-\frac{\hbar f_2}{\lambda}\phi_2(x) & &+ \hbar g_2 f_2 \rho_2(x)  &
+ \frac{\hbar\xi g_2}{\mu}\phi_2^\dagger(x) \cr
\quad & \quad & \quad & \quad \cr 
0 & \hbar\lambda g_1\phi_1(x) & 0
& -\hbar g_1^2  \rho_1(x) \cr
& +\frac{\hbar g_2}{\lambda}\phi_2(x) & &- \hbar g_2^2\rho_2(x) } \, .
$$
In the expression (\ref{c5}),
the sign of normal arrangement of operator factors is taken as
$\vdots\vdots $. The sign $\vdots\vdots$ , applied to the product of
several operator factors (including $\phi_1$,$\, \phi_2$,$ \, \phi_1^\dagger$
 and $\phi_2^\dagger$), ensures
the arrangement of all $\phi_1^\dagger$, $\phi_2^\dagger$ on the left, 
and all $\phi_1$, $\phi_2$ on the
right, {\it without altering the order
 of the remaining factors}.  For example,
$$ \vdots X \phi_1 \phi_2\phi_1^\dagger \phi_2^\dagger Y \vdots = \phi_1^
\dagger \phi_2^\dagger X Y \phi_1 \phi_2 \, ,$$
where $X$ and $Y$ may in general be taken as some functions of the
basic field operators.

Now one can easily check that  
  ${\cal L}( x; \lambda, \mu )$ (\ref{c6}) follows an equation given by
\beq
R( \lambda, \mu ){\cal L}( x; \lambda, \mu ) = 
{\cal L}( x; \mu, \lambda )R(\lambda, \mu) \, ,
\label{c7}
\eeq
where $R(\lambda, \mu)$ is a $(4\times 4)$  matrix of the form 
\beq
R( \lambda, \mu ) = \pmatrix { 1 & 0 & 0 & 0 \cr
0 & s( \lambda, \mu ) & t( \lambda, \mu ) & 0 \cr
0 & t( \lambda, \mu ) & s( \lambda, \mu ) & 0 \cr
0 & 0 & 0 & 1} \, ,
\label{c8}
\eeq
with $t(\lambda, \mu)=\frac{\lambda^2 - \mu^2}{\lambda^2 q - \mu^2
q^-1} , ~ s( \lambda, \mu ) = \frac{( q - q^-1 )\lambda\mu}
{\lambda^2 q - \mu^2 q^-1} $ and $ q = e^{-i\alpha} $.
The above equation (\ref{c7}) enables 
us to determine the exact expressions of the parameters 
$f_1$, $f_2$, $g_1$, $g_2$, $\alpha$ in terms of the coupling
constant $\xi$. We obtain :
\bea
~~~~~~~~~~~~~~~~~~~~~~~\hbar\xi= -\sin\alpha , 
~~~~ f_1 = g_2 = \frac{\xi e^{-i\alpha / 2}}{\cos \alpha / 2},~~~~
g_1 = f_2 = \frac{\xi e^{i\alpha / 2}}{\cos \alpha / 2} \, . \nn
~~~~~(3.9a,b,c)
\eea
\addtocounter{equation}{1}
Using eqns.(\ref{c5}) and (\ref{c7}), we find that 
 the monodromy matrix (\ref{c3}) satisfies QYBE given by
\beq
R( \lambda, \mu ){\cal T}^{x_2}_{x_1}( \lambda ) \otimes 
{\cal T}^{x_2}_{x_1}( \mu ) = {\cal T}^{x_2}_{x_1}( \mu ) \otimes 
{\cal T}^{x_2}_{x_1}( \lambda )R( \lambda, \mu ) \, .
\label{c10}
\eeq
Using the above QYBE (\ref{c10}), the commutation relations 
among all elements  of the quantum 
monodromy matrix (\ref {c3}) can be obtained easily. 

Eqns. (3.9a,b,c), describing the relations between $f_1, \, f_2, \, g_1,
\, g_2,  \, \alpha$ and the coupling
constant $\xi$, provide the necessary conditions for the Lax operator
(\ref{c1}) to satisfy QYBE (\ref {c10}). 
From eqn.(3.9a) we can conclude that, the above method 
of deriving QYBE for quantum BMT model is applicable only when the 
coupling constant
$\xi$ lies within the range $\v \xi \v \leq \frac{1}{\hbar}$.
The parameter $\alpha$ has a one-to-one correspondence with the
coupling constant $\xi$ for
$-\frac{\pi}{2} \leq \alpha \leq \frac{\pi}{2}$.  
For the purpose of investigating the classical limit of the quantum
Lax operator (\ref{c1}), we take the 
$\alpha \rightarrow 0$ limit which is equivalent to the $\hbar\rightarrow 0$ 
limit
for a fixed value of $\xi$. From eqns.(3.9b,c), it follows that at this limit
$f_1, f_2 \rightarrow \xi$ and $g_1, g_2 \rightarrow \xi$.  
Hence we find that the quantum Lax operator 
(\ref{c1}) correctly reproduces the classical Lax operator 
(2.1) at $\hbar \rightarrow 0$ limit.

\vspace{1cm}

\noindent \section {Algebraic Bethe ansatz for the quantum monodromy
matrix on an infinite interval}
\renewcommand{\theequation}{4.{\arabic{equation}}}
\setcounter{equation}{0}

\medskip

The quantum monodromy matrix in an infinite interval is
defined as  
\beq
{\cal T}(\lambda) = 
\lim_{\stackrel {x_2 \rightarrow + \infty} {x_1 \rightarrow -\infty}} 
e(-x_2,\lambda) {\cal T}^{x_2}_{x_1}(\lambda)e(x_1,\lambda) \, ,
\label {d1}
\eeq
where ${\cal T}^{x_2}_{x_1}(\lambda)$ is given by eqn.(\ref {c3}).
Just as in the classical case, the quantum Lax operator (\ref{c1}) also
satisfies the symmetry relations
\bea
~~~~~~~~~~~~~~~
{\cal U}_q( x,\lambda )^* = K  \, {\cal U}_q( x,\lambda^* ) \, K ,
\quad  {\cal U}_q( x, -\lambda ) = K^\prime  \,
{\cal U}_q(x,\lambda ) \,  K^\prime 
\, , \nn~~~~~~~~~~~~~~~~(4.2a,b)
\eea
\addtocounter{equation}{1}
where $K$ and $ K^\prime $ matrices have appeared earlier in eqn.(2.4).  
Using eqn.(4.2a), the quantum monodromy
matrix (\ref {d1}) can be expressed in a symmetric form given by
\bea
{\cal T}(\lambda)=\pmatrix {A(\lambda) & -\xi B^\dagger(\lambda) \cr
                          B(\lambda) & A^\dagger(\lambda)} \, ,
\label {d3}
\eea
where $\lambda$ is a real parameter.
From eqn.(4.2b), it follows that $A(-\lambda)= A(\lambda)$ and 
$B(-\lambda)= - B(\lambda)$. So it is sufficient to obtain
the commutation relations among the elements of
the quantum monodromy matrix (\ref {d3}) only for $\lambda \geq 0$.

Now we aim to obtain the infinite interval limit of the QYBE 
satisfied by ${\cal T}(\lambda)$ (\ref {d3}).
To this end, we split the 
${\cal L}( x; \lambda, \mu ) $ matrix (\ref {c6}) into two parts:
$$
{\cal L}( x; \lambda, \mu ) = {\cal L}_0( \lambda, \mu ) + {\cal L}_1( x;
\lambda, \mu ) \, ,
$$
where 
$ {\cal L}_0( \lambda, \mu ) $ is given by
$$
{\cal L}_0( \lambda, \mu ) = \lim_{|x| \rightarrow \infty} {\cal L}( x;
\lambda, \mu ) = \pmatrix {-\frac{i}{4}( \lambda^2 + \mu^2 ) & 0 & 0 & 0 \cr
+\frac{i}{4}(\frac{1}{\lambda^2} + \frac{1}{\mu^2}) & & &  \cr
&  &  &  \cr
0 & -\frac{i}{4}( \lambda^2 - \mu^2 ) & 0 & 0 \cr
& +\frac{i}{4}(\frac{1}{\lambda^2} - \frac{1}{\mu^2}) & & \cr
&  &  &  \cr
0 & -\hbar \xi \lambda \mu -\frac{\hbar\xi}{\lambda\mu} & 
\frac{i}{4}( \lambda^2 - \mu^2 ) & 0 \cr
&  & -\frac{i}{4}(\frac{1}{\lambda^2} - \frac{1}{\mu^2}) & \cr
&  &   &  \cr
0 & 0 & 0 & \frac{i}{4}( \lambda^2 + \mu^2 ) \cr
&  &  & -\frac{i}{4}(\frac{1}{\lambda^2} + \frac{1}{\mu^2})} \, ,
$$
and 
$ {\cal L}_1( x; \lambda, \mu ) $  is the field dependent part of
$ {\cal L}( x; \lambda, \mu ) $, which vanishes at $x\rightarrow \pm \infty$.
From eqn.(\ref{c7}) we get
\beq
R( \lambda, \mu ) \varepsilon( x; \lambda, \mu ) = \varepsilon( x; \mu,
\lambda ) R( \lambda, \mu ) \, ,
\label{d4}
\eeq
where $ \varepsilon( x; \lambda, \mu ) = e^{{\cal L}_0( \lambda, \mu )x}$.
By using the above mentioned splitting of $ {\cal L}( x; \lambda, \mu ) $,
 we derive the integral form of differential equation (\ref{c5}) as

$$
{\cal T}^{x_2}_{x_1}( \lambda ) \otimes {\cal T}^{x_2}_{x_1}( \mu ) = 
\varepsilon( x_2 - x_1; \lambda, \mu ) + \int_{x_1}^{x_2} dx \,
\varepsilon( x_2 - x; \lambda, \mu ) \, \vdots {\cal L}_1( x, \lambda,\mu )
{\cal T}^{x}_{x_1}( \lambda ) \otimes {\cal T}^{x}_{x_1}( \mu ) \vdots
\, .  
$$
From this integral relation it is clear that
 at the asymptotic limit $ x_1, x_2 \rightarrow \pm\infty $,
 $~{\cal T}^{x_2}_{x_1}( \lambda ) \otimes {\cal T}^{x_2}_{x_1}( \mu )
\rightarrow \varepsilon( x_2 - x_1; \lambda, \mu )$, which is an 
oscillatory term.  To get rid of this problem, we define an operator like   
\beq
W( \lambda, \mu ) = \lim_{\stackrel
{x_2 \rightarrow +\infty}{x_1 \rightarrow -\infty}}
\varepsilon( - x_2; \lambda, \mu ) 
{\cal T}^{x_2}_{x_1}( \lambda ) \otimes {\cal T}^{x_2}_{x_1}( \mu )
\varepsilon( x_1; \lambda, \mu ) \,.
\label{d5}
\eeq
In the above defined operator, the oscillatory nature of 
${\cal T}^{x_2}_{x_1}( \lambda ) \otimes {\cal T}^{x_2}_{x_1}( \mu )$ 
has been removed and $W(\lambda, \mu)$ is perfectly well behaved 
at the limit $ x_1, x_2 \rightarrow \pm\infty $.
By using (\ref{c10}) and (\ref{d4}), it is easy to verify that 
the operator $W( \lambda, \mu )$ (\ref {d5}) satisfies an equation given by
\beq
R( \lambda, \mu ) W( \lambda, \mu ) = W( \mu, \lambda ) R( \lambda, \mu ) \, .
\label{d6}
\eeq
The above equation represents the QYBE of BMT model at 
an infinite interval limit. 

Next, we want to express the QYBE (\ref {d6}) directly in terms of 
the monodromy matrices (\ref{d1}) defined in an infinite interval. 
For this purpose,  $W(\lambda, \mu)$ (\ref {d5}) can be rewritten as
\beq
W( \lambda, \mu ) 
= C_+( \lambda, \mu ) {\cal T}( \lambda ) \otimes {\cal T}(\mu)
 C_-( \lambda, \mu ) \, ,
\label{d7}
\eeq
where
\bea 
C_+( \lambda, \mu ) = \lim_{x \rightarrow \infty} \varepsilon( -x ;\lambda,
\mu ) E( x; \lambda, \mu ) , ~~~ C_-( \lambda, \mu ) =  \lim_{x \rightarrow
-\infty} E( -x; \lambda, \mu ) \varepsilon( x; \lambda,\mu ) \, ,  
~~ (4.8a,b) \nn
\eea
\addtocounter{equation}{1}
with $ E( x; \lambda, \mu ) = e( x, \lambda )\otimes e( x, \mu ) $.
Substituting the explicit  forms of $ E( x; \lambda, \mu ) $  and 
$\varepsilon(x; \lambda,\mu )$ to (4.8a,b),
 and taking the limits in the principal value sense:  
$\lim_{x \rightarrow \pm \infty}  P( \frac{e^{ikx}}{k} ) = \pm i \pi
\delta(k) $, we obtain   
\bea
C_+( \lambda, \mu ) = \pmatrix { 1 & 0 & 0 & 0 \cr
0 & 1 & 0 & 0 \cr
0 & \rho_+( \lambda, \mu ) & 1 & 0 \cr
0 & 0 & 0 & 1} \, ,~~~~
C_-( \lambda, \mu ) = \pmatrix { 1 & 0 & 0 & 0 \cr
0 & 1 & 0 & 0 \cr
0 & \rho_-( \lambda, \mu ) & 1 & 0 \cr                                 
0 & 0 & 0 & 1}  \, ,
\label {d10}
\eea
where
\bea
\rho_\pm( \lambda, \mu ) &=& \mp \frac{2i\hbar\xi \left(\lambda \mu +
\frac{1}{\lambda\mu}\right)}{\lambda^2 - \mu^2 -\frac{1}{\lambda^2} +
\frac{1}{\mu^2}}
+ 2 \pi \hbar \xi  
\left( \lambda \mu + \frac{1}{\lambda\mu}\right)\delta( \lambda^2 
- \mu^2-\frac{1}{\lambda^2}+\frac{1}{\mu^2}) \nn \\ 
&=& \mp \frac{2 i \hbar\xi 
\{\lambda \mu + \frac{1}{\lambda\mu}\}}{\lambda^2 - \mu^2-\frac{1}{\lambda^2}
+ \frac{1}{\mu^2} \mp i\epsilon} \, .
\label{rho}
\eea
Substituting the expression of $W(\lambda,\mu) $  (\ref{d7})
in eqn.(\ref {d6}), we can express this QYBE for the infinite interval in
the form 
\beq
R( \lambda, \mu ) C_+( \lambda, \mu ) {\cal T}( \lambda ) 
\otimes {\cal T}( \mu ) C_-( \lambda, \mu ) = 
C_+( \mu, \lambda ) {\cal T}( \mu ) \otimes {\cal T}(\lambda)
 C_-( \mu, \lambda ) R( \lambda, \mu ) \, .
\label{d11}
\eeq 

By inserting the explicit forms of $R(\lambda , \mu)$ (\ref {c8}),
 $C_\pm (\lambda , \mu )$  (\ref{d10}), and 
${\cal T}( \lambda )$ (\ref {d3}) to the above QYBE (\ref{d11}) and 
comparing the matrix elements from both sides of this equation, we obtain
the following commutation relations:
\bea
&&\left[ A( \lambda ), A( \mu ) \right] = 0 \, , ~~~~
\left[ A( \lambda ), A^\dagger( \mu ) \right] = 0 \, , ~~~~ 
\left[ B( \lambda ), B( \mu ) \right] = 0 \, , \nn 
~~~~~~~~~~~~~~~(4.12a,b,c) \\
&&A(\lambda) B^\dagger(\mu) = \frac{\mu^2 q - \lambda^2 q^{-1}}{\mu^2 -
\lambda^2 - i \epsilon} B^\dagger( \mu ) A( \lambda ) \nn \\
&&~~~~~~~~~~~~~~=  \frac{\mu^2 q - \lambda^2 q^{-1}}{\mu^2 -\lambda^2} 
B^\dagger( \mu ) A( \lambda ) - 2 \pi \hbar\xi \lambda \mu \delta( \lambda^2 -
\mu^2 ) B^\dagger( \lambda ) A( \mu ) \, , \nn 
~~~~~~~~~(4.12d) \\
&&B( \mu ) A( \lambda ) = \frac{\mu^2 q - \lambda^2 q^{-1}}{\mu^2 -\lambda^2 
- i \epsilon } A( \lambda ) B( \mu ) \nn \\
&&~~~~~~~~~~~~~
=  \frac{\mu^2 q - \lambda^2 q^{-1}}{\mu^2 -\lambda^2} A( \lambda ) B( \mu )
- 2 \pi \hbar \xi \lambda \mu \delta( \lambda^2 - \mu^2 ) 
A( \mu ) B( \lambda)
\, , \nn ~~~~~~~~~~~~~(4.12e) \\
&&B( \mu ) B^\dagger( \lambda ) = \tau(\lambda,\mu) 
 B^\dagger( \lambda ) B( \mu ) + 4 \pi\hbar\xi  \lambda \mu \delta(
\lambda^2 - \mu^2 ) A^\dagger( \lambda ) A( \lambda ) \, , \nn 
~~~~~~~~~~~~~~~~~(4.12f)
\eea
\addtocounter{equation}{1}  
where 
$$\tau( \lambda, \mu ) = \left[ 1 + \frac{8 \hbar^2\xi^2  \lambda^2
\mu^2}{{( \lambda^2 - \mu^2 )}^2} - \frac{4 \hbar^2\xi^2  {\{\lambda \mu
+ \frac{1}{\lambda\mu}\}}^2 }
{( \lambda^2 - \mu^2 - \frac{1}{\lambda^2} + \frac{1}{\mu^2} - i \epsilon ) 
( \lambda^2 - \mu^2 -\frac{1}{\lambda^2} + \frac{1}{\mu^2} + i \epsilon)} 
\right].$$ 
It is interesting to note that, for the case
$\lambda \neq \mu$, eqn.(4.12f) gives
$[ B(\lambda) , B^\dagger (\mu) ] \neq 0$, 
whereas from eqn.(2.16f), one obtains that
$ \{ b( \lambda ) , b^*( \mu ) \} =0$ for 
$\lambda \neq \mu$. Thus the correspondence principle  
is not manifest here in a
straightforward manner. However the $\hbar \rightarrow 0$ limit
of $\tau(\lambda, \mu) $, gives the correct classical 
counterpart of the commutation relation (4.12f).   

Due to eqn.(4.12a), all the operator valued coefficients occuring in the
expansion of $\ln A( \lambda)$ will commute among themselves. As a
consequence the BMT model described by the Lax operator (\ref{c1}) 
turn out to be a quantum integrable system. 
By applying the method of algebraic Bethe ansatz, one can also 
construct the exact eigenstates for all commuting operators which 
are generated through the expansion of $\ln A( \lambda)$.
With the help of eqn.(\ref {d1}), it is easy to find that
 $A(\lambda) \v 0 \r = \v 0 \r$.  By using this relation
and eqn.(4.12d), it can be shown that  
\bea
A(\lambda ) \,
 \v \mu_1, \mu_2, \cdots , \mu_N \r  
= \prod_{r=1}^N \left( { \mu_r^2 q - \lambda^2 q^{-1} \over
 \mu_r^2  - \lambda^2 -i \epsilon } \right)\, 
 \v \mu_1, \mu_2, \cdots, \mu_N\r \, ,
\label{d13}
\eea
where $\mu_j$s are all distinct real or complex numbers
 and $\v \mu_1, \mu_2, \cdots,
\mu_N \r  \equiv B^\dagger( \mu_1 ) B^\dagger( \mu_2 ) \cdots \\ B^\dagger(
\mu_N ) \v 0 \r $ represents a Bethe eigenstate. 
Using the commutation relation (4.12f)
one can also calculate the norm
of the eigenstates 
 $B^\dagger(\mu_1) B^\dagger(\mu_2) \cdots B^\dagger(\mu_N) \v 0 \r $.
However, the commutation relation 
(4.12f) contains product of singular functions $( \lambda^2
- \mu^2 - \frac{1}{\lambda^2} + \frac{1}{\mu^2} - i \epsilon )^{-1}( \lambda^2 
- \mu^2 - \frac{1}{\lambda^2} + \frac{1}{\mu^2}+ i \epsilon )^{-1} $, 
which is undefined at the limit $\lambda \rightarrow \mu$. 
 As a result, eigenstates like 
 $B^\dagger(\mu_1) B^\dagger(\mu_2) \cdots B^\dagger(\mu_N) \v 0 \r $ 
are not normalised on the $\delta$-function. To solve this problem,  
we consider a reflection operator given by
\beq 
R^\dagger(\lambda ) ~=~ B^\dagger(\lambda){(A^\dagger(\lambda))}^{-1}
\label{d22}
\eeq
and its adjoint $R(\lambda ) $. By using eqns.(4.12a-f), 
we find that such reflection operators satisfy well defined 
commutation relations like 
\bea
&&R^{\dagger}(\lambda)R^{\dagger}(\mu) 
= S^{-1}(\lambda , \mu) \, R^{\dagger}(\mu) R^{\dagger}(\lambda) \, , \nn \\
&&R(\lambda)R(\mu) = S^{-1}(\lambda , \mu) \, R(\mu) R(\lambda) \, , \nn \\
&&R(\lambda)R^{\dagger}(\mu) 
= S(\lambda , \mu) \, R^{\dagger}(\mu) R(\lambda) + 
4\pi\hbar \lambda^2 \delta(\lambda^2 - \mu^2) \, ,
\label {d23}
\eea
where 
\beq
S(\lambda , \mu) \, = \,  { {\lambda^2 q - \mu^2 q^{-1}} \over 
{\lambda^2 q^{-1} - \mu^2 q }} \,. 
\label{d24}
\eeq
The $S(\lambda , \mu) $ defined above represents the nontrivial
 $S$-matrix element of two-body scattering among the 
related quasi-particles.  We find that this 
$S(\lambda , \mu) $ satisfies the following conditions:
\beq 
S^{-1}(\lambda , \mu) = S(\mu ,\lambda) = S^*(\lambda , \mu) \, ,
\label {d25}
\eeq
and remains nonsingular at the limit $\lambda \rightarrow \mu$.  
Consequently,  the action of the operators like 
$R^\dagger(\lambda)$ on the vacuum would produce well defined states which  
can be normalised on the $\delta$-function. 

The point to be noted here is that in eqn.(\ref{d13}), the eigenvalues of
$A(\lambda)$  are in general complex. To get real
eigenvalues, we define a new operator 
$\ln {\hat A}(\lambda)$ through the relation 
$ \ln {\hat A}(\lambda) \equiv 
 \ln A(\lambda  e^{{-i \alpha\over 2}}) $ and expand this operator
in inverse powers of $\lambda $:
\beq
 \ln {\hat A}(\lambda) = 
\sum_{n=0}^{\infty} \frac{i{\cal C}_n}{\lambda^{2n}}  \, .
\label {d14}
\eeq
Using eqns.(\ref {d13}) and (\ref {d14}), it is easy to see that  
${\cal C}_n$s satisfy eigenvalue equations like 
$$
{\cal C}_n \, \v \mu_1, \mu_2, \cdots, \mu_N\r \, , =  \chi_n \,
\v \mu_1, \mu_2, \cdots, \mu_N\r \, ,
$$
 where the first few $\chi_n$s are  explicitly given by
\beq 
\chi_0 = \alpha N \, , ~~~~ \chi_1 = 2 \sin \alpha \sum_{j=1}^N \mu_j^2 \, ,~~~
\chi_2 = \sin 2\alpha \sum_{j=1}^N \mu_j^4 \, .
\label {d15}
\eeq
It may be noted that these eigenvalues are all real when $\mu_j$s are 
taken as real numbers.
Next we expand the operator $\ln {\hat A}(\lambda)$ in powers of 
$\lambda $ as 
\beq
\ln {\hat A}(\lambda) = 
\sum_{n=0}^{\infty} i\tilde{\cal C}_n \lambda^{2n}  \, ,
\label {d16}
\eeq
and by using (\ref {d13}) we obtain 
$$
{\tilde {\cal C}}_n \, \v \mu_1, \mu_2, \cdots, \mu_N\r \, , = 
 {\tilde \chi}_n \, \v \mu_1, \mu_2, \cdots, \mu_N\r \, .
$$
The first few ${\tilde \chi}_n$s are explicitly given by
\beq
{\tilde \chi}_0 = -\alpha N \, ,
~~~~ {\tilde \chi}_1 = -2 \sin \alpha \sum_{j=1}^N 
\frac{1}{\mu_j^2} \, , ~~~~
{\tilde \chi}_2 = -\sin 2\alpha \sum_{j=1}^N \frac{1}{\mu_j^4} \, .
\label {d17}
\eeq
In analogy with the classical case,
one can now define the momentum and  
Hamiltonian of the quantum BMT model as 
$${\cal P} = -\frac{1}{4\xi}(C_1 + \tilde{C}_1),~~~~~
{\cal H}= -\frac{1}{4\xi}\left( C_1 - \tilde{C}_1 \right). $$ 
By using (\ref{d15}) and (\ref{d17}),
the eigenvalue
equations corresponding to the above momentum and Hamiltonian are obtained
as
\bea 
&& {\cal P}\v \mu_1, \mu_2, \cdots, \mu_N \r = \frac{1}{2}
\sum_{j=1}^N \left( \mu_j^2 -
\frac{1}{\mu_j^2} \right)\v \mu_1, \mu_2, \cdots, \mu_N \r \, ,\nn \\
&&{\cal H}\v \mu_1, \mu_2, \cdots, \mu_N \r = \frac{1}{2}
\sum_{j=1}^N \left( \mu_j^2 +
\frac{1}{\mu_j^2} \right)\v \mu_1, \mu_2, \cdots, \mu_N \r \, .
\label{d18}
\eea
In the above expressions, $\mu_j$s are taken as real numbers and
$\v \mu_1, \mu_2, \cdots, \mu_N \r$ represents a scattering state. Now to
construct quantum $N$-soliton states of BMT model, complex values of $\mu_j$
can be chosen in such a way so that the eigenvalues corresponding to
different expansion coefficients of $\ln {\hat A(\lambda)}$ 
still remains real. Such a choice is given by 
\bea
\mu_j= \mu \exp\Big[-i\alpha\left(\frac{N+1}{2}-j\right)\Big] \, ,  
\label{d19}
\eea
where $\mu$ is a real parameter and $j \in [1,2, \cdots ,N]$. For the
above choice of $\mu_j$, eqn.(\ref{d13}) takes the form
\bea
A(\lambda)\v \mu_1, \mu_2, \cdots , \mu_N \r = q^{-N}\Big( \frac
{\lambda^2 - \mu^2 q^{N+1}}{\lambda^2 - \mu^2 q^{-N+1}} \Big)
\v \mu_1, \mu_2, \cdots , \mu_N \r \, .
\label{d20}
\eea
Consequently, the energy eigenvalue equation corresponding to the quantum 
$N$-soliton state can be obtained as
${\cal H}\v \mu_1, \mu_2, \cdots , \mu_N \r = E 
\v \mu_1, \mu_2, \cdots , \mu_N \r \, $, where
\bea
E =\frac{1}{2}
\left( \mu^2 + \frac{1}{\mu^2}
\right) \frac{\sin \alpha N}{\sin \alpha} .
\label{d21}
\eea
Thus we find that 
quantum N-soliton states can be constructed for BMT model 
for $N>1$.
Now we assume a particular value of the coupling constant $\xi$ given by
$\xi = -\sin \alpha = - \sin(\frac{2\pi m}{n})$, where $m$ and $n$ are nonzero
integers which do not have any common factor. From eqn.(\ref{d19}), we
obtain $\mu_j = \mu_{j+n}$ for the above choice of $\xi$. Since all the
$\mu_j$s have to be distinct, we get $ N\leq n$ as a 
restriction on the
number of quasi-particles that can form a quantum soliton state for BMT model
when $\xi = - \sin(\frac{2\pi m}{n})$.

Next we aim to calculate the binding energy for a $N$-soliton state of
quantum BMT model. Substituting the expression of $\mu_j$ (\ref{d19}) to
the first relation in eqn.(\ref{d18}), 
the momentum eigenvalue of a $N$-soliton state is obtained as
\beq
P = \frac{1}{2} (\mu^2 -\frac{1}{\mu^2})
\frac{\sin \alpha N}{\sin \alpha}
\, .
\label{eigen}
\eeq
It is intersting to observe that the energy (\ref{d21}) and the momentum
eigenvalue (\ref{eigen}) of a $N$-soliton state satisfy the dispersion 
relation $E^2 = P^2 + m^2 $,
where $m= \frac{\sin\alpha N}{\sin \alpha}$.
To calculate binding energy we assume that the momentum $P$ (\ref{eigen}) is
equally distributed among $N$ number of single-particle scattering states.
The real wave number associated with each of these single particle states is
denoted by $\mu_0$.
With the help of eqns.(\ref{d18}) and (\ref{eigen}), we find that
\bea
\mu_0^2 - \frac{1}{\mu_0^2} = (\mu^2 -\frac{1}{\mu^2})
\frac{\sin \alpha N}{N\sin \alpha}\, .
\label{bin1}
\eea
Using eqn.(\ref{d18}), the total energy for $N$ number of such single particle
states is obtained as
\bea
E^\prime = \frac{N}{2}\left(\mu_0^2 + \frac{1}{\mu_0^2}\right) =
\frac{N}{2}{\left\{{\left(\mu^2-\frac{1}{\mu^2}\right)}^2 \frac{\sin^2
\alpha N}{N^2\sin^2
\alpha} + 4\right\}}^{\frac{1}{2}}\, .
\label{bin2}
\eea

Subtracting $E$ (\ref{d21}) from $E^\prime$ (\ref{bin2}), we obtain the binding
energy of the quantum $N$-soliton state as
\bea
E_B(\alpha, N) &=& E^\prime - E \nn \\
&=&\frac{N}{2}{\left\{{\left(\mu^2-\frac{1}{\mu^2}\right)}^2 
\frac{\sin^2\alpha N}{N^2\sin^2
\alpha} + 4\right\}}^{\frac{1}{2}} - \frac{1}{2} \left(\mu^2 +\frac{1}
{\mu^2}\right)\frac{\sin \alpha N}{\sin \alpha}\, .   
\label{bin3}
\eea
Note that the above expression of $E_B(\alpha, N)$ remains invariant
under the transformation $\alpha \rightarrow -\alpha$. So it is sufficient 
to analyse the nature of binding energy within the range 
$0 < \alpha \leq \frac{\pi}{2} $.
Now, for $E_B(\alpha,N)$ to represent the energy of a real bound state, 
 $E^\prime$ has to be greater than $E$. Since $E^\prime$ (\ref{bin2}) is
always positive, it is evident that $E^\prime > E $ for $E< 0$. So we will
restrict our attention only for the case $E>0$, when the condition $E^\prime
> E$ is equivalent to ${E^\prime}^2 > E^2$.   
Substituting the explicit expressions for $E^\prime$ (\ref{bin2})
and $E$ (\ref{d21}), the above condition takes the form
\bea
N\sin \alpha > \sin \alpha N \, .
\label{bin4}
\eea
Substituting $N=2$ in (\ref{bin4}), we get the trivial inequality
$1 > \cos \alpha$ for $\alpha>0$. 
So the condition (\ref{bin4}) is satisfied for $N=2$ case within 
our chosen range of $\alpha$. 
By using the method of
induction, we can easily prove that the condition (\ref{bin4}) is  
valid for arbitrary values of $N$. Thus we get an $N$-soliton bound state 
when $\alpha$ lies in the range $0 < \v \alpha \v \leq \frac{\pi}{2} $.
%To use the method of induction, let us first assume
%that
%\bea
%N \sin \alpha > \sin \alpha N \, ,
%\label{bin5}
%\eea
%when $\alpha$ lies in the range $0 < \alpha \leq \frac{\pi}{2} $.
%Next we consider an identity given by
%\bea
%\sin \alpha (N+1) = \sin \alpha N \cos \alpha + \cos \alpha N \sin \alpha \,.
%\label{bin6}
%\eea
%Using the inequality (\ref{bin5}) and the above identity, we obtain
%\bea
%\sin \alpha (N+1) < \sin \alpha ( N \cos \alpha + \cos \alpha N)\, .
%\label{bin7}
%\eea
%Since for $0 < \alpha \leq \frac{\pi}{2},~ N\cos \alpha < N$ and 
%$\cos \alpha N \leq 1$, we obtain
%\bea
%\sin \alpha (N+1) < (N+1) \sin \alpha \, .
%\label{bin8}
%\eea
%Thus by assuming $N\sin \alpha > \sin \alpha N$ for 
%$0 < \alpha \leq \frac{\pi}{2} $, we are able to show that \\
%$(N+1)\sin \alpha > \sin \alpha (N+1)$ for the same range of $\alpha$. We
%have also observed earlier that this inequality holds true for the 
%$N=2$ case for the above range of $\alpha$.  
%Hence by using the method of induction, it is established that ${E^\prime}^2
%> E^2$ for $0<\alpha \leq \frac{\pi}{2}$ and hence $E^\prime > E$ 
%within this range of $\alpha$. 

\medskip

\noindent \section {Concluding Remarks}
In this article we consider the classical Lax opeartor of BMT model 
and obtain the PB relations
among various elements of the classical monodromy matrix at the infinite
interval limit. 
By using these PB relations,
the classical integrability of BMT model is established in the 
Liouville sense. 
We also calculate the classical conserved quantities of BMT model.
Next, we quantise the Lax opeartor of BMT model. By using a variant of
QISM, that can be directly applied to the field
theoretic models, we obtain the QYBE for the quantum monodromy matrix at a 
finite interval. This QYBE enables us to determine 
the various parameters of the quantum Lax operator in terms of the
coupling constant $\xi$. Then we take the infinite
interval limit of this QYBE and 
derive all possible commutation
relations among the various elements of the corresponding quantum monodromy
matrix. These commutation relations enable us to establish the quantum
integrability of BMT model and also to construct the exact eigenstates for
the quantum version of the Hamiltonian (\ref{a1}) as well as other 
conserved quantities by using algebraic Bethe
ansatz. We also obtain the commutation relation between creation and
annihilation operators associated with quasi-particles of BMT model and find
out the $S$-matrix for two body scattering.

In this context, we consider the BMT model with some special values
of coupling constant given by $\xi = -\sin \alpha = -\sin(\frac{2\pi
m}{n})$, where $m$ and $n$ are nonzero integers with no common factor. 
It turns out that the number of quasi-particles, which form a bound state
for such quantum BMT model, cannot exceed the value of $n$. We have also
derived the exact expression of binding energy for a $N$-soliton state of 
quantum BMT model. The binding energy turns out to be positive for  
all allowed values of $\alpha$.

The commutation relation between creation and annihilation operators   
will play an important role in the future study, since by using it one might
be able to calculate the norm of Bethe eigenstates and various correlation
functions of the BMT model. In future, we would also like to obtain 
the quantum conserved quantities of BMT model in terms of the field
operators by using a method which was used earlier in the case of
nonlinear Schrodinger model [17] and DNLS model [16]. 

\medskip

\noindent {\bf Acknowledgments }

The author would like to thank Dr. B. Basu-Mallick for many 
valuable suggestions and careful reading of the manuscript.

\newpage

\noindent {\bf Appendix A}

\medskip

Here we give the details of deriving eqn.(\ref{c5}).
Direct attempt to calculate \\ 
$\frac{\partial}{\partial x_2}\left({\cal T}^{x_2}_{x_1}(\lambda)\otimes 
{\cal T}^{x_2}_{x_1}(\mu)\right)$ by using eqn.(\ref {c4}), 
leads to indeterminate 
expressions of the form $\left[ {\cal T}^{x_2}_{x_1}
(\lambda), \phi_1^\dagger( x_2 ) \right]$ and $\left[ {\cal T}^{x_2}_{x_1}
(\lambda), \phi_2^\dagger( x_2 ) \right]$.
To avoid this problem by using the method of extension [3], 
we shift the upper limit of the monodromy 
matrix ${\cal T}^{x_2}_{x_1}(\lambda) $
by a small amount $\epsilon $ and take  $\epsilon 
\rightarrow 0$ limit only after differentiating the product 
${\cal T}^{x_2 +\epsilon }_{x_1}(\lambda)\otimes {\cal T}^{x_2}_{x_1}(\mu)$
with respect to $x_2$.  
So, using eqn.(\ref {c4}), we obtain
\bea
\frac{\partial}{\partial x_2} 
\left ( \, {\cal T}^{x_2 + \epsilon}_{x_1}( \lambda ) \otimes 
{\cal T}^{x_2}_{x_1}( \mu ) \, \right)
 &=&   \vdots \,  \left( \, {\cal U}_q( x_2 + \epsilon; \lambda )
\otimes \one + \one \otimes {\cal U}_q( x_2; \mu ) \, \right)
{\cal T}^{x_2 + \epsilon}_{x_1}( \lambda ) \otimes 
{\cal T}^{x_2}_{x_1}( \mu ) \, \vdots  \nn \\
&~&+ \,  K_+ \, + \,  K_- ~ ,  
~~~~~~~~~~~~~~~~~~~~~~~~~~~~~~~~~~~~~~~~~~~~~~~~(A1) \nn
\eea 
where
\bea
K_+ &=& i \xi \mu \left[ {\cal T}^{x_2 + \epsilon}_{x_1}( \lambda ),
\phi_1^\dagger( x_2 ) \right] \otimes \sigma_+ {\cal T}^{x_2}_{x_1}( \mu ) 
-\frac{i \xi}{\mu }\left[ {\cal T}^{x_2 + \epsilon}_{x_1}( \lambda ),
\phi_2^\dagger( x_2 ) \right] \otimes \sigma_+ {\cal T}^{x_2}_{x_1}( \mu )
\nn \\
&+& i f_1 \left[ {\cal T}^{x_2 + \epsilon}_{x_1}( \lambda ),
\phi_1^\dagger( x_2 ) 
\right] \otimes 
e_{11} {\cal T}^{x_2}_{x_1}( \mu ) \phi_1( x_2 ) \nn \\ 
&-&i f_2 \left[ {\cal T}^{x_2 + \epsilon}_{x_1}( \lambda ),\phi_2^\dagger( x_2 ) 
\right] \otimes 
e_{11} {\cal T}^{x_2}_{x_1}( \mu ) \phi_2( x_2 ) \nn \\
&-& i g_1 \left[ {\cal T}^{x_2 + \epsilon}_{x_1}( \lambda ),
\phi_1^\dagger( x_2 ) 
\right] \otimes e_{22} {\cal T}^{x_2}_{x_1}( \mu ) \phi_1(x_2) \nn \\ 
&+& i g_2 \left[ {\cal T}^{x_2 + \epsilon}_{x_1}( \lambda ),\phi_2^\dagger( x_2 ) 
\right] \otimes e_{22} {\cal T}^{x_2}_{x_1}( \mu ) \phi_2(x_2) \, , 
\nn \\
K_- &=& i \lambda \sigma_- {\cal T}^{x_2 + \epsilon}_{x_1}( \lambda ) \otimes 
\left[ \phi_1( x_2 + \epsilon ), {\cal T}^{x_2}_{x_1}( \mu ) \right] 
- \frac{i}{\lambda}\sigma_- {\cal T}^{x_2 + \epsilon}_{x_1}( \lambda ) \otimes 
\left[ \phi_2( x_2 + \epsilon ), {\cal T}^{x_2}_{x_1}( \mu ) \right] \nn \\
&+& i f_1 \phi_1^\dagger( x_2 + \epsilon )e_{11} 
{\cal T}^{x_2 + \epsilon}_{x_1}( \lambda ) \otimes 
\left[ \phi_1( x_2 + \epsilon ), {\cal T}^{x_2}_{x_1}( \mu ) \right] \nn \\ 
&-& i f_2 \phi_2^\dagger( x_2 + \epsilon )e_{11} 
{\cal T}^{x_2 + \epsilon}_{x_1}( \lambda ) \otimes 
\left[ \phi_2( x_2 + \epsilon ), {\cal T}^{x_2}_{x_1}( \mu ) \right] \nn \\
&-& i g_1 \phi_1^\dagger( x_2 + \epsilon ) e_{22}{\cal T}^{x_2 + \epsilon}_{x_1}
( \lambda ) \otimes \left[ \phi_1( x_2 + \epsilon ), 
{\cal T}^{x_2}_{x_1}( \mu ) \right] \nn \\ 
&+& i g_2 \phi_2^\dagger( x_2 + \epsilon ) e_{22}{\cal T}^{x_2 + \epsilon}_{x_1}
( \lambda ) \otimes \left[ \phi_2( x_2 + \epsilon ), 
{\cal T}^{x_2}_{x_1}( \mu ) \right] \, .\nn
\eea

Now we consider the case, $\epsilon > 0$. 
Since $\phi_1( x_2 + \epsilon)$ and $\phi_2( x_2 + \epsilon )$ commute with 
$\phi_1( x ), \, \phi_1^\dagger( x ), \, \phi_2( x), \, \phi_2^\dagger (x) $ 
for all $x$ lying within $x_1$ and $x_2$,  we get
$\left[ \phi_1( x_2 + \epsilon ), {\cal T}^{x_2}_{x_1}( \mu ) \right] = 
\left[ \phi_2( x_2 + \epsilon ), {\cal T}^{x_2}_{x_1}( \mu ) \right] =0$.
Thus we can conclude that for a positive $\epsilon $,  
$K_- = 0 $ . So we have to calculate only the
nontrivial commutators $\left[ {\cal T}^{x_2 + \epsilon}_{x_1}
( \lambda ), \phi_1^\dagger( x_2 ) \right]$  and 
$\left[ {\cal T}^{x_2 + \epsilon}_{x_1}
( \lambda ), \phi_2^\dagger( x_2 ) \right]$ appearing in the expression of
$K_+$.  

First let us calculate the commutator $\left[ {\cal T}^{x_2 + \epsilon}_{x_1}
( \lambda ), \phi_1^\dagger( x_2 ) \right]$. For this purpose, we 
consider a `transformation' $\Omega$, which replaces the classical variables 
$\psi(x)$ and $\psi^*(x)$ by quantum operators $\psi(x)$ and
$\psi^\dagger(x)$ respectively. 
Next we use a correspondence principle [3], 
\bea
~~~~~~~~~~~~~~\left[ {\cal T}^{x_2 + \epsilon}_{x_1}( \lambda ), 
\phi_1^\dagger( x_2 ) \right]
= i\hbar  : \Omega\left\{ T^{x_2 + \epsilon}_{x_1}( q; \lambda ), 
\phi_1^*( x_2 ) \right\} : \, , \nn 
~~~~~~~~~~~~~~~~~~~~~~~~~~~~~~ (A2)
\eea 
where
$ T^{x_2 + \epsilon}_{x_1}( q; \lambda )$  represents a classical 
monodromy matrix given by
$$
T^{x_2 + \epsilon}_{x_1}( q; \lambda ) =
{\cal P} \exp \int_{x_1}^{x_2} 
U_q(x,\lambda) dx   \, ,
$$
and $U_q( x, \lambda ) = {\Omega}^{-1}{\cal U}_q( x, \lambda )$.  
By using the fundamental PB relations
(\ref{a2}), it is easy to find that
\bea
\{ T^{x_2 + \epsilon}_{x_1}( q; \lambda ), \phi_1^*( x_2 ) \} &=& 
\int^{x_2 + \epsilon}_{x_1} dx~ T^{x_2 + \epsilon}_x( q; \lambda ) 
\{ U_q( x, \lambda ), \phi_1^*( x_2 ) \} T^x_{x_1}( q, \lambda ) \nn \\
&=& T^{x_2 + \epsilon}_{x_2}( q; \lambda )( f_1 \phi_1^*( x_2 ) e_{11} - g_1
\phi_1^*( x_2 ) e_{22} + \lambda \sigma_- ) T^{x_2}_{x_1}( q; \lambda ) \, . \nn
\eea  
Taking $\epsilon 
\rightarrow 0$ limit of the above expression and substituting it in (A.2),  
we obtain
\bea
~~~~~~\lim_{\epsilon \rightarrow 0} \,
\left[ \, {\cal T}^{x_2 + \epsilon}_{x_1}( \lambda )
, \phi_1^\dagger( x_2 ) \, \right] = i \hbar \left( f_1 \phi_1^\dagger( x_2 ) 
e_{11} - g_1
\phi_1^\dagger( x_2 ) e_{22} + \lambda \sigma_- \right) \, {\cal T}^{x_2}_{x_1}
(\lambda) \, . \nn 
~~~~~~~~~~(A3)
\eea

Next we have to calculate the commutator $\left[ \, {\cal T}^
{x_2 + \epsilon}_{x_1}( \lambda )
, \phi_2^\dagger( x_2 ) \, \right]$. Using the same correspondence principle
as before
and finally taking the $\epsilon \rightarrow 0 $ limit one obtains,
\bea
~~~~~~\lim_{\epsilon \rightarrow 0} \,
\left[ \, {\cal T}^{x_2 + \epsilon}_{x_1}( \lambda )
, \phi_2^\dagger( x_2 ) \, \right] = i \hbar \left( - f_2 \phi_2^\dagger( x_2 ) 
e_{11} + g_2
\phi_2^\dagger( x_2 ) e_{22} -\frac{1}{\lambda} \sigma_- \right) \, {\cal T}^{x_2}_{x_1}
(\lambda) \, . \nn 
~~~~~~~(A4)
\eea 

Taking the $\epsilon \rightarrow 0$ limit of eqn.(A1) and 
using (A3) and (A4), we finally obtain the differential 
equation (\ref{c5}). Note that, instead of $\epsilon >0$,
 we could have chosen $ \epsilon < 0 $ in eqn.(A1). In that case only 
the commutators 
$\left[ \phi_1( x_2 + \epsilon ), {\cal T}^{x_2}_{x_1}( \mu ) \right]$ and
$\left[ \phi_2( x_2 + \epsilon ), {\cal T}^{x_2}_{x_1}( \mu ) \right]$
give
nontrivial contributions.
However,by repeating similar steps as outlined above and finally taking 
the $\epsilon \rightarrow 0$ limit, 
we would have obtained the same differential equation (\ref{c5}).

\newpage
\leftline {\large \bf References}
\medskip
\begin{enumerate}
\item L. D. Faddeev, Sov. Sci. Rev. C1 (1980) 107. 

\item L. D. Faddeev, in: J. B. Zuber, R. Stora (Eds.), Recent Advances in
Field Theory and Statistical Mechanics, 
North-Holland, Amsterdam, 1984, p.561.

\item  E. K. Sklyanin, in: M. Jimbo (Ed.),  
Yang-Baxter Equation in Integrable systems, Adv. Ser. in Math. Phys. 
Vol. 10, World Scientific, Singapore, 1990, p.121.

\item V. E. Korepin, N. M. Bogoliubov and A. G. Izergin, {\it Quantum Inverse 
Scattering Method and Correlation Functions} (Cambridge Univ. Press, 
Cambridge, 1993) and references therein. 

\item  Z. N. C. Ha, {\it Quantum Many-Body Systems in One Dimension} 
(World Scientific, Singapore, 1996) and references therein.

\item P. P. Kulish and E. K. Sklyanin, in: J. Hietarinta et al (Ed.), Lecture 
notes in Physics, Vol. 151, Springer Verlag, Berlin, 1982, p.61.  

\item M. Wadati, T. Deguchi and Y. Akutsu, Phys. Rep. 180 (1989) 247.

\item M. Takahashi, {\it Thermodynamics of One-Dimensional Solvable Models} 
(Cambridge Univ. Press, Cambridge, 1999).

\item A. Kl\"{u}mper and C. Scheeren in: A. Kundu (Ed.), Classical and
Quantum Nonlinear Integrable System : Theory and Applications, Inst. of
Physics, 2003.

\item F. G\"{o}hmann and V. E. Korepin, Phys. Lett. A 263 (1999) 293.

\item F. Haldane, J. Phys. C: Solid State Phys 14 (1981) 2585.

\item S. J. Tans et al, Nature 386 (1997) 474.

\item B. Basu-Mallick and A. Kundu, Phys. Lett. B 287 (1992) 149.

\item A. Kundu and B. Basu-Mallick, J. Math. Phys. 34 (1993) 1052.

\item B. Basu-Mallick and T. Bhattacharyya, Nucl. Phys. B 634 [FS] (2002)
611.

\item B. Basu-Mallick and T. Bhattacharyya, Nucl. Phys. B 668 (2003) 415.

\item K. M. Case, J. Math. Phys. 25 (1984) 2306.

\end{enumerate}

\end{document}